\documentclass[twocolumn,prb,showpacs,multicol,amsmath,amssymb]{revtex4}
\usepackage[dvips]{graphicx}
\usepackage{graphicx}
\usepackage{dcolumn}
\usepackage{bm}
\usepackage{graphics}
\usepackage{epsfig,color}

\newcommand{\be}{\begin{equation}}
\newcommand{\ee}{\end{equation}}
\newcommand{\bea}{\begin{eqnarray}}
\newcommand{\eea}{\end{eqnarray}}

\newcommand{\bS}{\bf S}

\begin{document}
\title{ First-order commensurate-incommensurate magnetic phase transition in the coupled FM spin-1/2 two-leg ladders}
\author{ J. Jahangiri, H. Hadipour, S. Mahdavifar, and S. Farjami Shayesteh}
\affiliation{Department of Physics, University of Guilan, 41335-1914, Rasht, Iran}

\affiliation{}
\date{\today}


\begin{abstract}
  We consider the spin-1/2 two-leg ladders with ferromagnetic (FM) interactions along legs and rungs. Using the stochastic series expansion QMC method, we study the low-temperature magnetic behavior of the system. An isolated spin-1/2 FM two-leg ladder is in the gapped saturated FM phase at zero temperature.  As soon as the spin-1/2 FM two-leg  ladders are connected with antiferromagnetic (AFM) inter-ladder interaction, a first-order commensurate-incommensurate quantum phase transition occurs in the ground state magnetic phase diagram. In fact a jump in the magnetization curve is observed. We found that, coupled  spin-1/2 FM two-leg  ladders are in  a nonmagnetic phase at zero temperature. Applying a magnetic field, the ground state of coupled spin-1/2 FM two-leg  ladders remains in the nonmagnetic phase up to a quantum saturate critical field.
 \end{abstract}
\pacs{75.10.Jm; 75.10.Pq}

\maketitle


\section{Introduction}\label{sec1}

Recently, spin-1/2 two-leg ladder systems have devoted  considerable growth to  itself experimentally and theoretically\cite{Dagotto,Uehara,Fujiwara,Vanishri,Plekhanov}. The main interest to study of these systems is related to this fact that the high-Tc superconductivity \cite{Uehara,Fujiwara,Vanishri} phenomenon occurs in these systems and also they have a gap in the spin excitation spectrum.

In the study of spin-1/2 two-leg ladder systems with antiferromagnetic (AFM) leg and rung interactions, the formation of spin singlets located on each rung open the spin gap in the energy spectrum which is called the gaped spin liquid phase\cite{Dagotto, Rieira}.  These kind of AFM two-leg ladder systems are observed in the nature\cite{Azuma, Ruegg,Watson, Hong, Oosawa, Giamarchi, Millet,Elfimov,Korotin, McCarron,Siegrist,Roth,McElfresh,Japaridze1, Jahangiri1}. The effect of a magnetic field on the physical properties of these compounds has
been a field of intense studies. It is found that there are two quantum critical fields in the ground state phase diagram of these kind of spin-1/2 two-leg ladders. Generally, at low magnetic fields ($h < h_{c_1}$),  there is a spin liquid phase (a gapped phase) at low temperature \cite{Dagotto}. Both magnetic
susceptibility and the magnetization go up first with cooling, then decay exponentially to zero at low temperatures. Also, the specific
heat has a single peak at low temperature due to transition from disordered phase to the spin singlet gapped phase.\cite{Ruegg,Wang}. The Tomonaga-Luttinger liquid (TLL) gapless phase is found in $h_{c_1} < h <
h_{c_2}$ regime at low temperatures\cite{Ruegg,Wang}.
 One of the spin liquid ($h <
(h_{c_1} + h_{c_2})/2$) or spin polarized ($h > (h_{c_1} +
h_{c_2})/2$) phases at higher temperature is expected. The
thermodynamic properties like magnetization and the susceptibility
have a finite value at low temperature which show the vanishing of
the energy gap in the TLL phase. Specific heat shows a second peak
and goes down linearly with lowering temperature in the TLL regime \cite{Ruegg}.

The spin-1/2 two-leg ladder systems with AFM legs and FM rungs are also observed from experimental point of view\cite{Masuda05, Tsujii09}. By means of the specific heat and the magnetocaloric effect measurements, a phase boundary between the spin liquid phase and the ordered phase is determined\cite{Tsujii09}. We have to mention that, the ladders with FM leg and AFM rung exchange interactions found experimentally during last two years\cite{Yamaguchi13, Yamaguchi14, Yamaguchi}.

Recently, Nagashiwa and coworker synthesized chlorido-bridged dinuclearcopper(II) complex with 2-methylisothiazol-3(2H)-one  with chemical formula [{Cu$^{(II)}$ Cl(O-mi)}$_{2}$(μ-Cl)$_{2}$)]\cite{Nagasawa}. This compound is a quasi-two-leg ladder system with FM exchange interaction in legs and rungs. The Weiss temperature is estimated about 8.7 K, indicating ferromagnetic behavior. Also no exotic phenomena due to spin frustration were observed within the measured temperature range. From theoretical point of view, spin-1/2 two-leg ladders with FM legs and rungs are much less studied. In a very recent work, the temperature dependence of the magnetic susceptibility and the magnetic structure factors is studied using the modified spin wave theory and the numerical exact diagonalization technique\cite{Hida12}. They have showed that in an intermediate temperature range, their analytical results are consistent with the numerical exact diagonalization results. The need for investigation of AFM inter-ladder coupling in FM ladders ( FM exchange interaction in both legs and rungs ) is also motivated by synthesizing the 3-Cl-4-F-V, 3-Br-4-F-V and 3-I-V crystals recently which are the candidates
for two-leg ladders with FM legs and AFM rungs \cite{Yamaguchi13, Yamaguchi14, Yamaguchi}.

In this paper we study the AFM inter-ladder interaction effect between FM two-leg ladders in the presence of a magnetic field (see Fig. 1). This system for large inter-ladder interaction can be considered as the ladders with FM legs and AFM rungs by FM coupling. We used the recently developed stochastic series expansion (SSE)\cite{Sandvik,Syljuasen} QMC method to provide numerical simulation results. The magnetization, the susceptibility and the specific heat are calculated for large enough finite size systems.  Our simulation results show that isolated FM two-leg ladders are in an ordered phase at zero temperature. As soon as the AFM inter-ladder interaction is added, a first-order commensurate-incommensurate quantum phase transition occurs from the gapped ferromagnetic phase to a nonmagnetic phase. By applying a uniform magnetic field, the ground state of coupled ladders remains in this  nonmagnetic phase up to a saturated critical field.

The outline of the paper is as follows: In section II we introduce our numerical simulation results for an isolated FM spin-1/2 two-leg ladder. In
section III we consider coupled FM ladders and present results of our QMC simulation on the low-temperature magnetic behavior of the system.  Finally we conclude and summarize our results in section IV.


\begin{figure}[t]
\centerline{\psfig{file=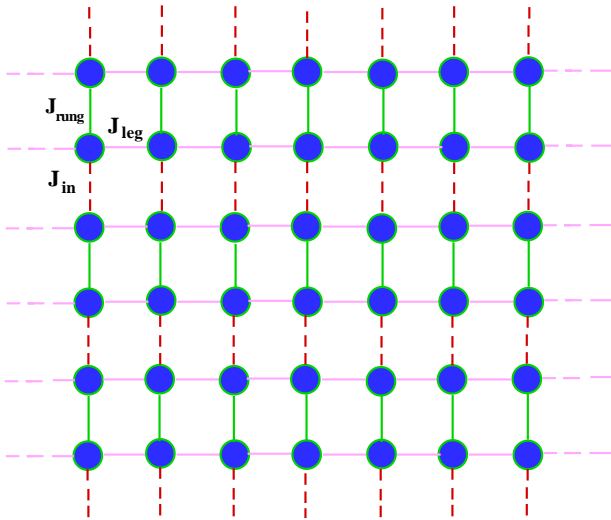,width=3.2 in}}
 \caption{The schematic picture of FM two-leg ladders with AFM inter-ladder interaction. The size of coupled ladders is $10\times30$. }\label{fig1}
\end{figure}
 \begin{figure}[t]
\centerline{\psfig{file=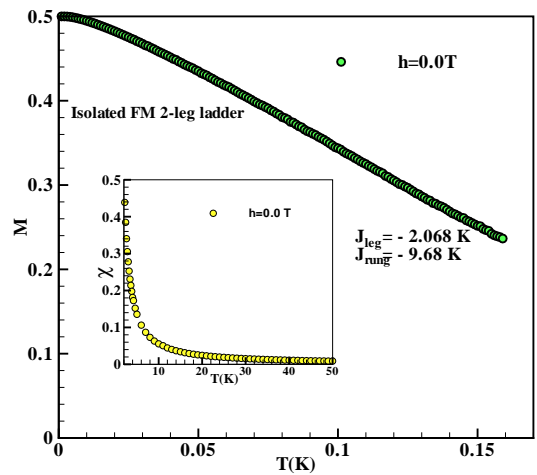,width=2.8 in}}
 \caption{The spontaneous magnetization as a function of the temperature for isolated spin-1/2 FM two-leg ladders. As it is seen, at zero temperature the ground state of the system is in a saturated FM phase. By increasing the temperature the induced thermal fluctuations decrease the magnetization of the system. In the inset, the magnetic susceptibility in the absence of the magnetic field is plotted as a function of the temperature. QMC simulation has been carried out for Heisenberg model. The size of two-leg ladder is $2 \times 100$. } \label{fig2}
\end{figure}


\section{An isolated FM two-leg ladder}\label{sec3}

In this section we consider a FM spin-1/2 two-leg ladder system in the presence of an external magnetic field. The Hamiltonian of the system is written as
\begin{eqnarray}
{H} &=& J_{leg} \sum_{n,\alpha} {\bS}_{n,\alpha} \cdot
{\bS}_{n+1,\alpha} - g \mu_{B} h  \sum_{n,\alpha}
S^{z}_{n,\alpha} \nonumber\\
&+& J_{rung} \sum_{n}{\bS}_{n,1} \cdot {\bS}_{n,2} \,  \label{Hamiltonian}
\end{eqnarray}
where $\bS_{n,\alpha}$ is the spin $S=1/2$ operator on rung $n$
($n=1,...,L$) and leg $\alpha$ ($\alpha=1,2$). An applied magnetic field $h$ in the $Z$ direction leads to zeeman term. The rung and leg exchanges are denoted by $J_{rung}$ and $J_{leg}$, respectively. \\
It is found\cite{Nagasawa} that the rung exchange $J_{rung}=-9.68~K$ is about four times larger than  $J_{leg}=-2.068~K$. In the following, we consider these values in our QMC simulation approach. We have used the ALPS\cite{Albu07} code which is known as one of the best codes in this field. We have considered two-leg ladders with the maximum size $N=2 L=2 \times 100$ and periodic boundary conditions. The QMC simulation is performed for the maximum $1000000$ equilibration sweeps and $2000000$ measurement steps.

 \begin{figure}[t]
\centerline{\psfig{file=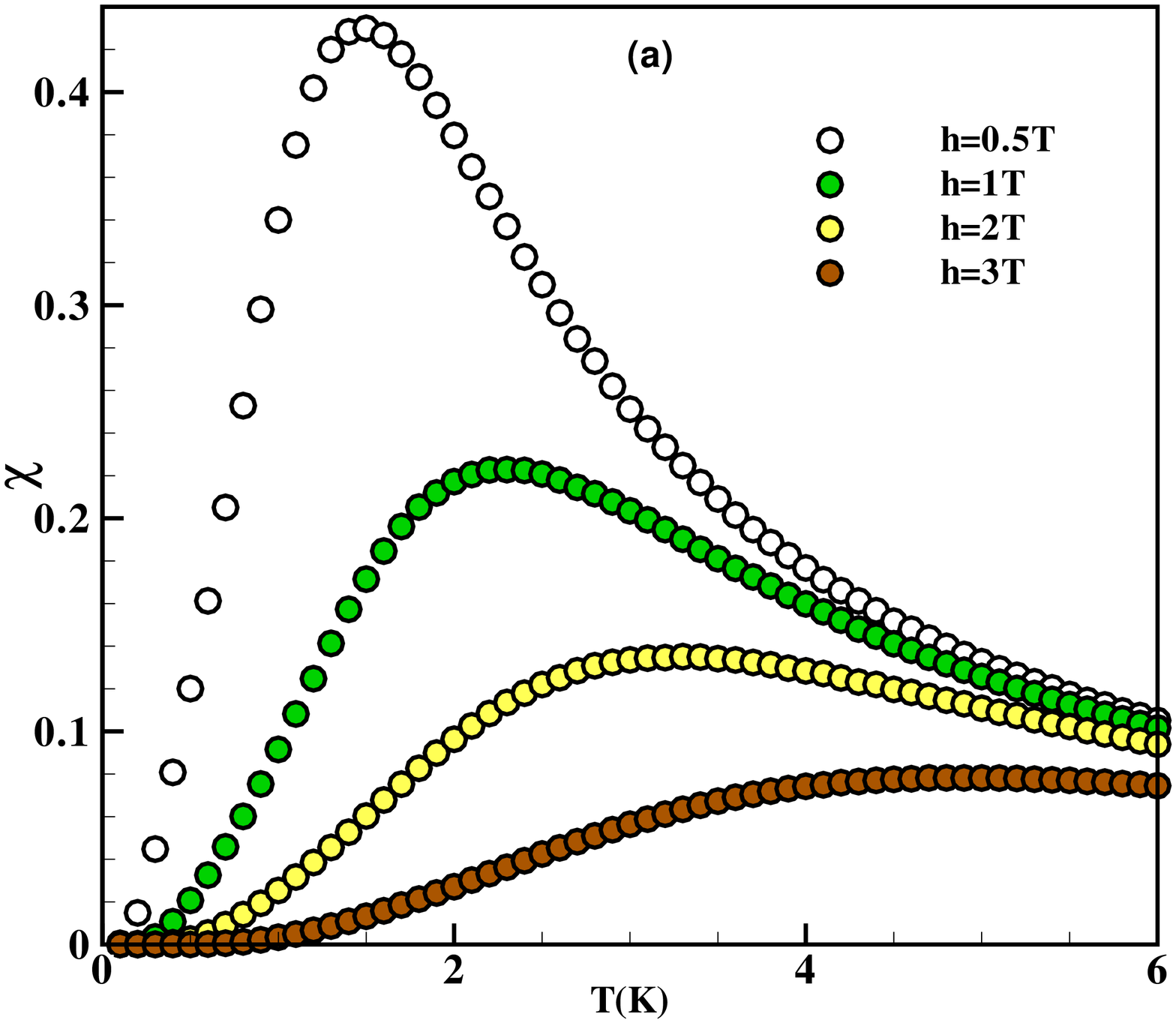,width=2.8 in}}
\centerline{\psfig{file=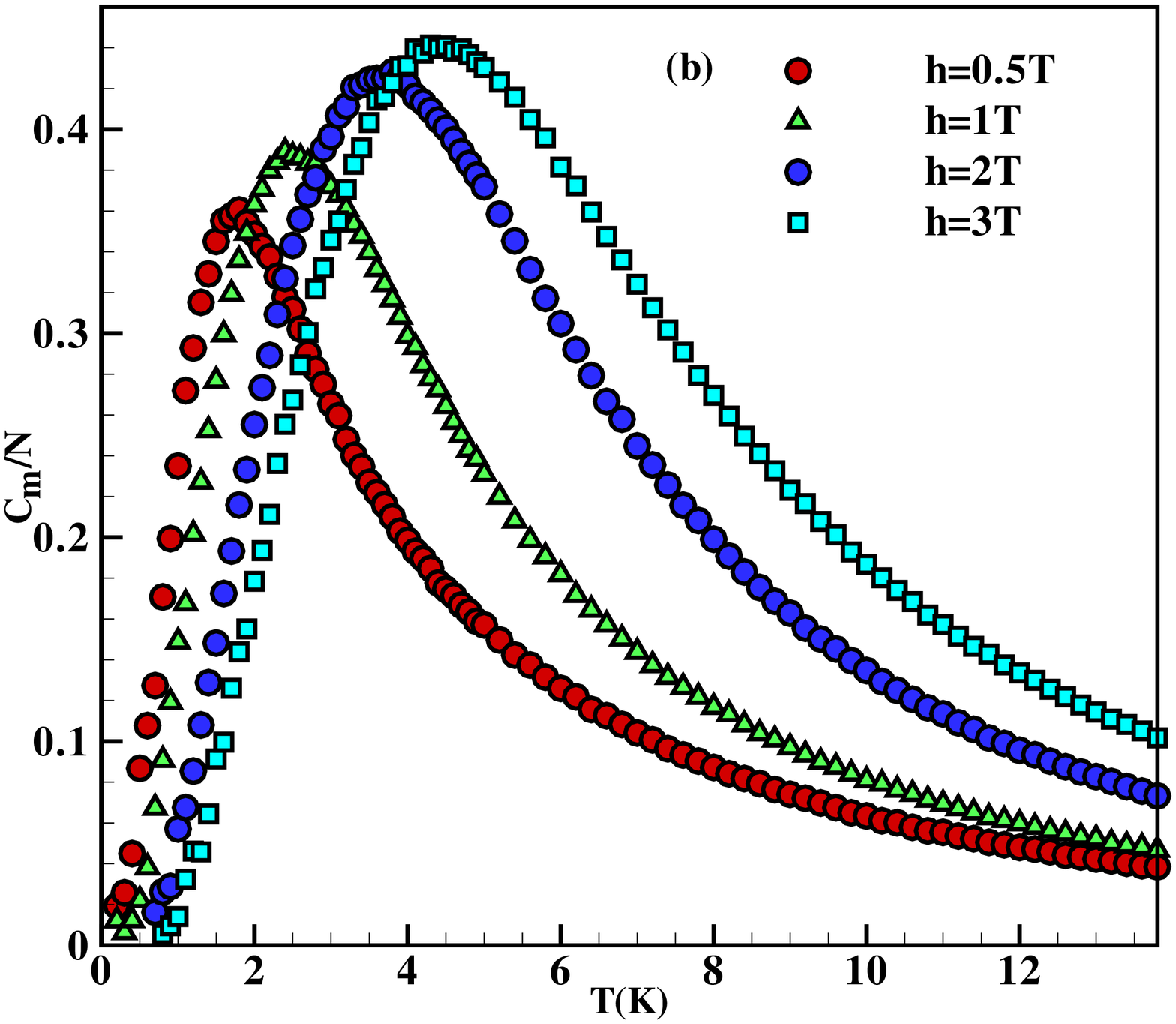,width=2.8 in}}
\caption{(a) Magnetic susceptibility $\chi(T)$ versus temperature for isolated spin-1/2 FM two-leg ladders at different values of the magnetic
field. QMC simulation has been carried out for Heisenberg model.(b) Magnetic specific heat per site versus the temperature for different values of the magnetic field. QMC simulation has been carried out for Heisenberg model. The size of two-leg ladder is $2 \times 100$. } \label{fig3}
\end{figure}

Fig.~\ref{fig2} presents the magnetization as a function of the temperature in the absence of the  magnetic field. Induced thermal fluctuations by increasing temperature cause in decreasing the magnetization and it reaches zero for high enough temperatures. The magnetic susceptibility is also shown in the inset of Fig.~\ref{fig2}. It only goes up with cooling the system which is in good agreement with the experimental results\cite{Nagasawa}. The asymptotic behavior confirms that the ground state of the system is in a magnetic ordered
phase at $T = 0$, which is also in complete agreement with our results on the magnetization.
 In addition, we have also studied the effect of the magnetic field on the thermodynamic behavior of an isolated FM  two-leg ladder system. Our QMC simulation results are presented in Fig.~\ref{fig3}(a) for different values of the magnetic field $h=0.5~T$, $1.0~T$, $2.0~T$, and $3.0~T$. we have calculated the spin gap according to expression $\chi(T)\sim$exp$(-\Delta/T)/\sqrt{T}$. The obtained results are $\Delta=0.88~K$, $1.6~K$, $2.6~K$, and $4.5~K$, for $h=0.5~T$, $1.0~T$, $2.0~T$, and $3.0~T$, respectively.\\
  As shown in Fig.~\ref{fig3}(a), the magnetic susceptibility goes up first with lowering temperature until it reaches to a maximum value, then decreases exponentially down to zero at low temperatures. This exponential fall of susceptibility indicates that there is a gap in the excitation spectrum of the FM  two-leg ladder system in the presence of a magnetic field.  With decreasing the magnetic field the peak of susceptibility curve increases and thus the spin gap is weakened.\\
   To complete our study on the thermodynamics of an isolated FM spin-1/2 two-leg ladder, we have also calculated the specific heat, $C_{m}(T)$. Our numerical results are presented in Fig.~\ref{fig3}(b) for different values of the magnetic field. As it is seen, at zero temperature the specific heat is zero, by increasing the temperature it remains almost zero up to a threshold temperature which is known as the indication of the spin gap. We have also checked the low temperature results and found an exponential behavior  which is also an indication of the gapped phase.

\section{Coupled FM two-leg ladder system}\label{sec3}

 High-field nuclear magnetic resonance and inelastic neutron scattering measurements in some quasi-two-leg ladder systems\cite{Oosawa, Giamarchi, Yamaguchi13, Yamaguchi14} show that weak
coupling between ladders induces $3D$ phase transition at low temperature region.  In these compounds, phase transition between quantum phases and  the 3D gapless-Neel ordered phase occurs at  very low temperatures.

To the best of our knowledge, the effect of the inter-ladder coupling in spin-1/2 two-leg ladders with
FM legs and rungs has not been studied so far. Interesting results are not expected for the inter-ladder FM coupling. For this reason, we have considered the inter-ladder AFM coupling. We have performed the QMC simulations on the AFM coupled FM two-leg ladder systems for different values of the inter-ladder exchange $J_{in}$ and in the presence of a magnetic field. The scheme of coupled ladders is illustrated in Fig.~\ref{fig1}. The numerical results are calculated for coupled ladders with $J_{rung}=-9.68~K$, $J_{leg}=-2.068~K$, $J_{in}/J_{leg}=-0.1, -0.2, ..., -0.5$ and different sizes up to $5\times2 L=5\times(2\times30)$.

\begin{figure}[t]
\centerline{\psfig{file=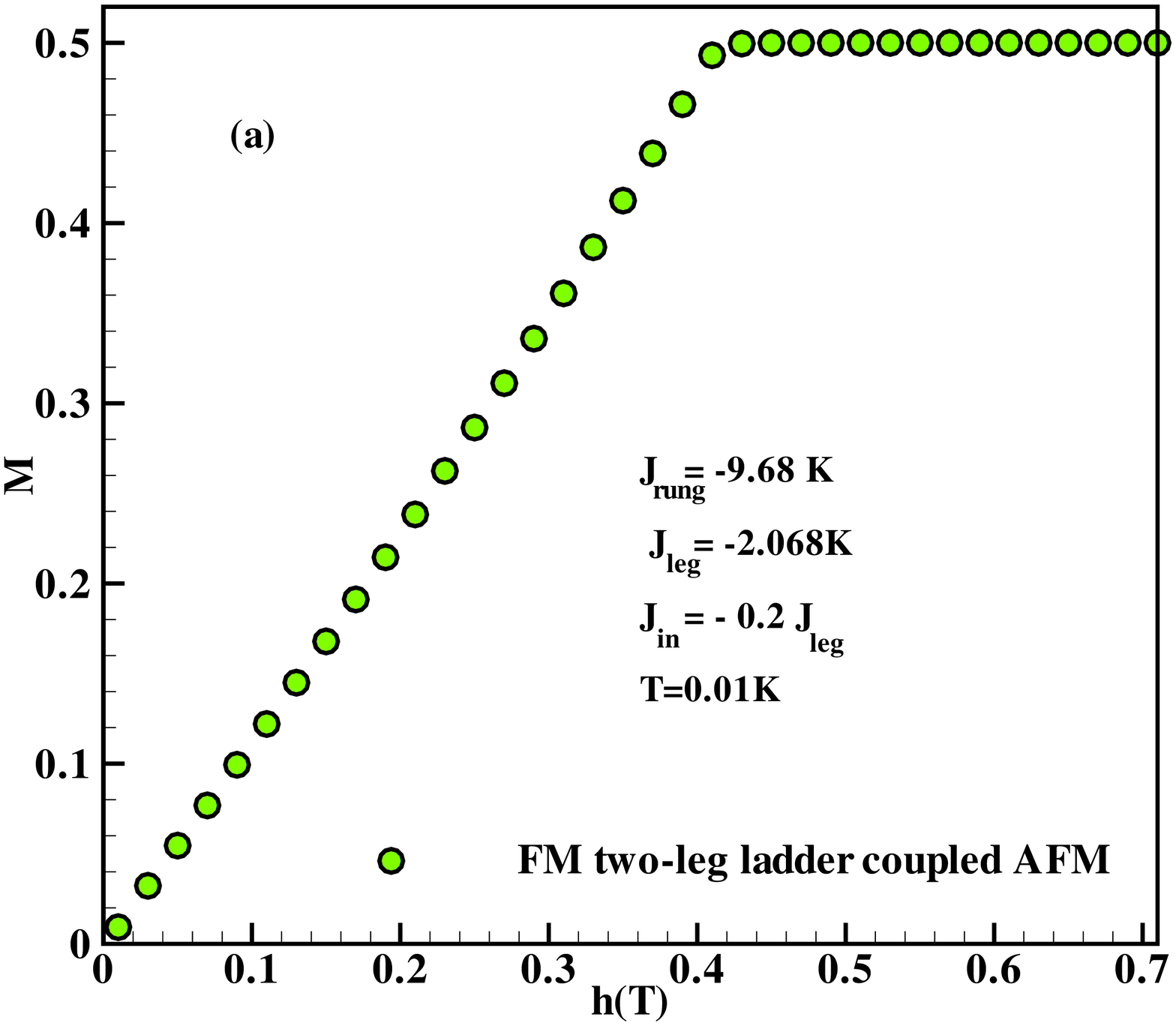,width=2.8 in}}
\centerline{\psfig{file=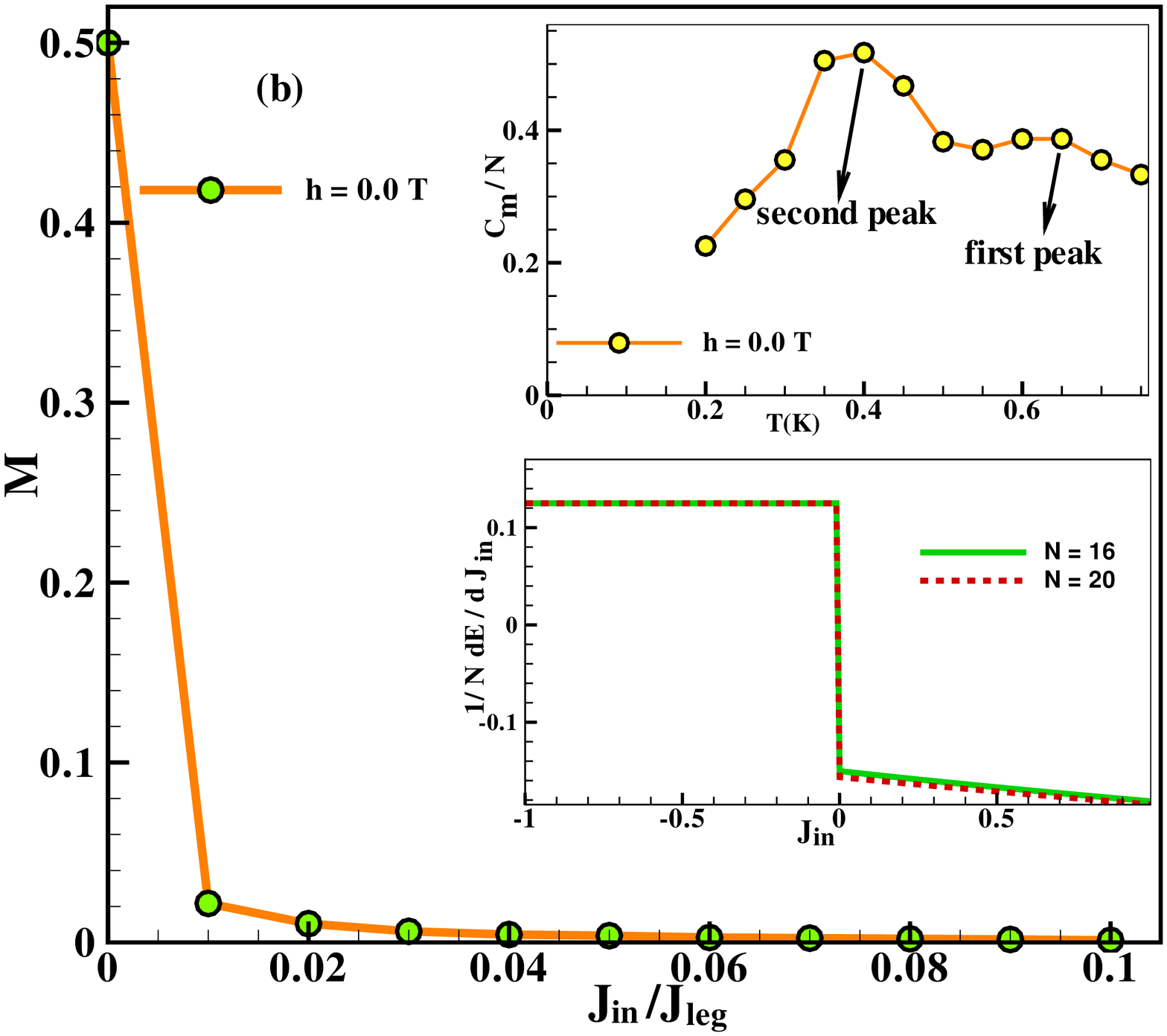,width=2.8 in}}
 \caption{(a) The magnetization of spin-1/2 FM two-leg ladders with inter-ladder AFM interaction as a function of the magnetic field in the vicinity of the zero temperature and $J_{in}/J_{leg}=-0.2$. QMC simulation has been
carried out for Heisenberg model. The size of coupled ladders is $10\times30$. (b) The magnetization versus inter-ladder $J_{in}$ interaction in the absence of the magnetic field. The size of coupled ladders is $10\times30$. The effect of AFM inter-ladder interaction causes the system undergoes a first-order phase transition. QMC calculation has been carried out for Heisenberg model. In the above inset, specific heat per site was plotted as a function of the temperature in the absence of the magnetic field by using QMC simulation. In the bottom inset, using numerical lancsoz algorithm, the first derivative of energy per site versus $J_{in}$ was plotted.} \label{fig4}
\end{figure}

Fig.~\ref{fig4}(a) shows the magnetization, $M$, of the coupled system as a function of the magnetic field in the vicinity of the zero temperature and $J_{in}/J_{leg}=-0.2$. It is clearly seen  that the magnetization is zero in the absence of the magnetic field  which shows that the induced quantum fluctuations by inter-ladder interaction destroy the long-range FM order of the ground state of isolated FM two-leg ladders. It means that in the ground state phase diagram of the coupled FM ladders, the line $J_{in}/J_{leg}=0$ must be a quantum critical line. By applying the magnetic field, the magnetization starts to increase and will be saturated in a critical magnetic field, which depends on the value of the inter-ladder exchange interaction.
So, such behavior of magnetization upon contribution of AFM inter-ladder interaction suggests, the
existence of a nonmagnetic phase.

Now let us see what happens when we attempt to change the inter-ladder interaction at zero temperature and in the absence of the magnetic field. Fig.~\ref{fig4}(b) presents the magnetization versus inter-ladder $J_{in}$ interaction in the absence of the magnetic field. The inter-ladder interaction effects embedded
in the coupled ladders drastically change the value of magnetization. There is a sharp drop of magnetization in the low inter-ladder coupling about $0.01J_{leg}$. This behavior is known as the main indication of a first-order phase transition. Therefore it suggests the existence of a first-order commensurate-incommensurate quantum phase transition in the ground state phase diagram of the model. Moreover, to confirm the type of the mentioned quantum phase transition in our model, we have also implemented the Lanczos algorithm to find the ground state energy of the system.   A very important indication of the first order phase transition is the discontinuity in the first derivative of the ground state energy at the quantum critical point. Using the numerical Lanczos method, we have calculated the ground state energy for system sizes $2\times(2\times L)=16, 20$ and plotted the first derivative of the ground state energy per site as a function of the inter-ladder interaction in the bottom inset of Fig. 4(b). The
results show a clear discontinuity in the first derivative of ground state energy which is in agreement with our QMC results.
  To find out the effect of inter-ladder AFM interaction, we have also presented the QMC results for the specific heat, $C_m$, in the top inset of Fig.~\ref{fig4}(b). In this figure, we observe a second peak at very low temperatures which can be considered as an indication of the mentioned nonmagnetic phase.

 \begin{figure}[t]
\centerline{\psfig{file=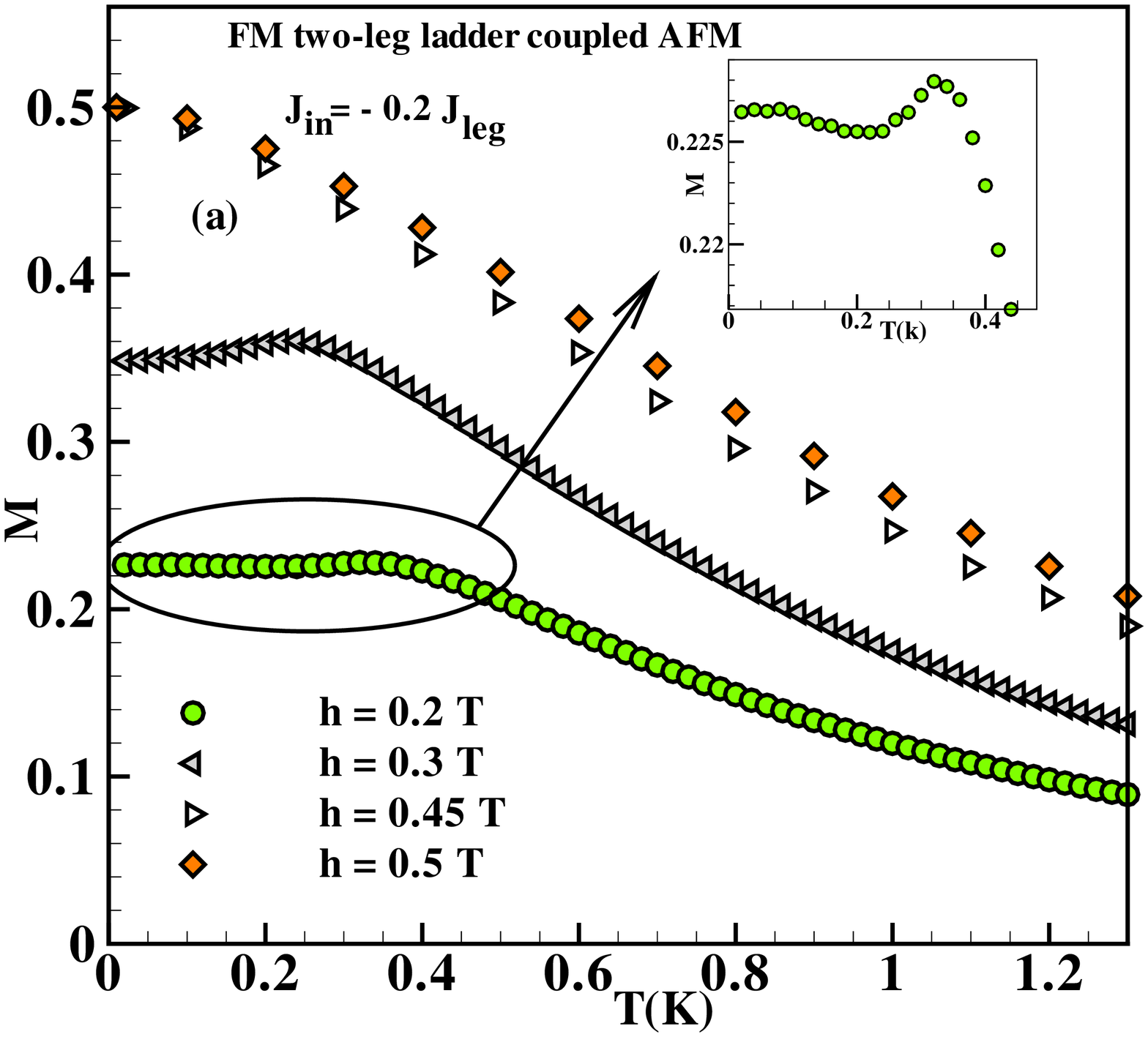,width=2.8 in}}
\centerline{\psfig{file=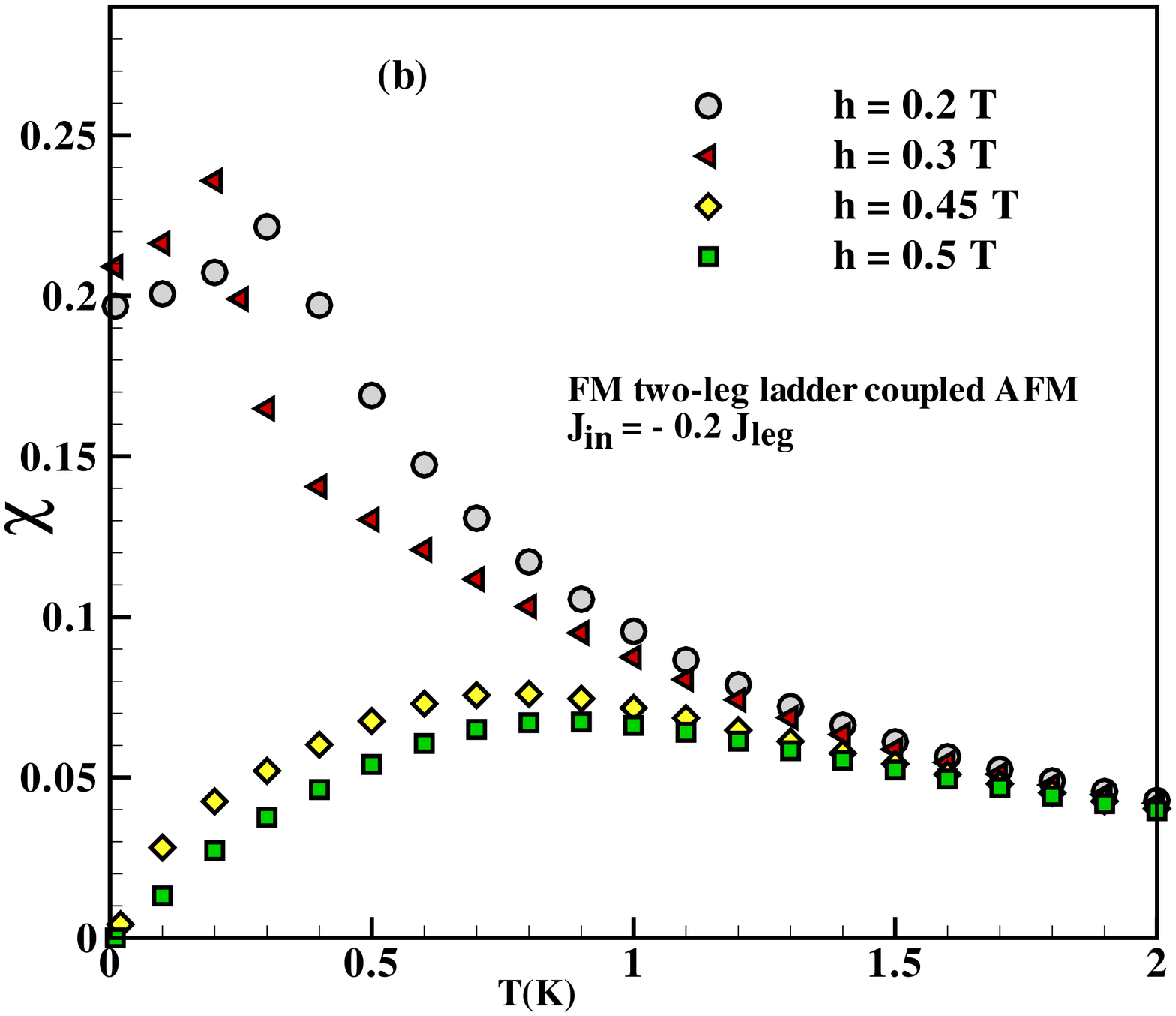,width=2.8 in}}
 \caption{(a) Magnetization versus temperature for spin-1/2 FM two-leg ladders coupled with AFM inter-ladder interaction at different values of the magnetic field. QMC simulation has been carried out for Heisenberg model. (b) the magnetic susceptibility versus the temperature for different values of the magnetic field above and below quantum critical field. QMC simulation has been carried out for Heisenberg model. The size of coupled ladders is $10\times30$.} \label{fig5}
\end{figure}


It is interesting to discuss the thermodynamic properties of coupled ladders in the gapless phase $J_{in}=-0.2~J_{leg}$ for different magnetic fields.
In general there is a similarity between isolated FM ladders and the AFM coupled FM ladders for high magnetic fields. But, the differences can be noticed within the range of $0.0<h<0.5~T$.
Fig.~\ref{fig5}(a) shows the magnetization versus the temperature for $h=0.2~T$, $h=0.3~T$, $h=0.45~T$, and $h=0.5~T$. The magnetization is not saturated for
$h<0.45~T$ at zero temperature. To find the critical magnetic field $h_{c}$, we have performed the calculation for other magnetic fields within the range of $0~T<h<0.5~T$.
There is no saturation in magnetization curve up to $h_{c}=0.42~T$ which is in accordance with Fig.~\ref{fig4}(a). In the inset of Fig.~\ref{fig5}(a) we have plotted the magnetization as a function of the temperature to show that there is not any plateau in the curve of the magnetization.  As shown in Fig.~\ref{fig5}(b) the magnetic susceptibility is plotted for different magnetic fields in the ratio of $J_{in}/J_{leg}=-0.2$. There are indications that the FM two-leg ladder spin systems under this weak AFM inter-ladder interaction undergo to a nonmagnetic phase in the low magnetic fields. $\chi(T)$ goes to  finite values at low temperatures within the range of $h<h_{c}$. Also this behavior suggests that the coupled ladders should be in gapless phase, whereas the exponential fall-off of $\chi(T)$ above critical field is a signature of existence of gap in this system.

\begin{figure}[t]
\centerline{\psfig{file=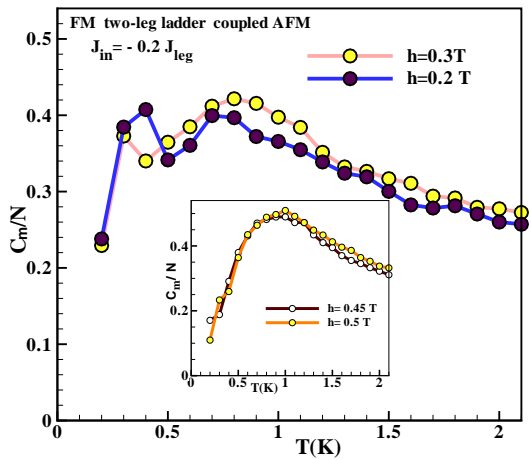,width=2.8 in}}
 \caption{ Magnetic specific  heat per site versus the temperature for spin-1/2 FM two-leg ladders coupled with AFM inter-ladder interaction in the region of the magnetic field $h<h_c=0.42T$. In the inset, the magnetic specific  heat per site is plotted for values of the field more than the quantum critical field. QMC simulation has been carried out for Heisenberg model. The size of coupled ladders is $10\times30$.  }\label{fig6}
\end{figure}

In Fig.~\ref{fig6} the curve of $C_{m}(T)$ at $h<h_{c}=0.42~T$ and $h>h_{c}$ (in the inset) is depicted.It can be clearly seen that  below the quantum critical field, a second peak appears at low temperatures, whereas as illustrated in the inset, above the quantum critical field $h_{c}$, the second peak disappears. These results indicate that the system undergoes a cross-over from gapped phase to a nonmagnetic phase. Also this behavior suggests that the coupled ladders should be in gapless phase. Generally, the weak AFM inter-ladder interaction has the substantial effect at low temperatures within the range of low magnetic fields $h<h_{c}=0.42~T$.


\section{Conclusion}\label{sec4}

In summary, we have calculated the thermodynamic properties of FM two-leg ladder like [{Cu$^{(II)}$ Cl(O-mi)}$_{2}$(μ-Cl)$_{2}$)]. We have performed stochastic series expansion QMC using ALPS code to investigate spin-1/2 isolated FM  two-leg ladders and the effect of the AFM inter-ladder exchange interaction on the low-temperature behavior of these type of ladders. For an isolated spin-1/2 FM two-leg ladder, a gapped behavior was observed at zero temperature. In principle, the ground state of an isolated spin-1/2 FM two-leg ladder is in the saturated FM phase.

As soon as the inter-ladder AFM exchange interaction is added, a first-order commensurate-incommensurate quantum phase transition occurs in the ground state phase diagram. Numerical results of magnetization and the specific heat showed that the gapped FM phase is replaced by a nonmagnetic phase. In the presence of a magnetic field, the ground state of the coupled ladders remains in this  nonmagnetic phase up to a critical field $h_c$. The value of the mentioned critical field depends on the value of the  inter-lader AFM interaction. Although in the case of [{Cu$^{(II)}$ Cl(O-mi)}$_{2}$(μ-Cl)$_{2}$)], the compound consists of isolated two-leg ladder, but such an enhancement of coupled interaction may occur upon chemical substitution.

Existence of the non-magnetic phase in this model is very similar to what can be probably observed in the coupled spin-1/2 two-leg ladders with FM
leg and AFM rung exchange interactions in the presence of a magnetic field\cite{Yamaguchi13, Yamaguchi14, Yamaguchi}.
In fact, two-leg ladders with FM leg and rung interactions consisting of weak AFM inter-ladder coupling can also be considered as the two-leg ladders
with FM leg and weak AFM rung exchange interactions which are FM coupled. The observation of non-magnetic phase in these new type of coupled ladders is stimulating for the future studies.


\section{Acknowledgement}\label{sec5}
We express our heartfelt thanks to G. I. Japaridze and T. Vekua for stimulating discussions.
Also, we thank Dr. Javad Vahedi for reading this manuscript.

\vspace{0.3cm}

\end{document}